\documentclass[prl, twocolumn,superscriptaddress,amsmath,amssymb,showpacs,floatfix,preprintnumbers]{revtex4-2}

\usepackage{amsmath}
\usepackage{graphicx}
\usepackage{dcolumn}
\usepackage{xcolor}
\usepackage{physics}
\usepackage{xr}
\usepackage{siunitx}
\usepackage{tabularx}
\usepackage{makecell}
\usepackage{multirow}
\DeclareGraphicsExtensions{.png .jpg .pdf}

\usepackage[normalem]{ulem}

\makeatletter
\newcommand*{\addFileDependency}[1]{
  \typeout{(#1)}
  \@addtofilelist{#1}
  \IfFileExists{#1}{}{\typeout{No file #1.}}
}
\makeatother
 
\newcommand*{\myexternaldocument}[1]{%
    \externaldocument{#1}%
    \addFileDependency{#1.tex}%
    \addFileDependency{#1.aux}%
}

\myexternaldocument{si}

\begin{document}

\title{Charge-to-spin conversion in twisted graphene/$\textrm{WSe}_2$ heterostructures}

\author{Seungjun Lee}\thanks{These authors contributed equally to this work.}
\author{D. J. P. de Sousa}\thanks{These authors contributed equally to this work.}
\affiliation{Department of Electrical and Computer Engineering, University of Minnesota, Minneapolis, Minnesota 55455, USA}

\author{Young-Kyun Kwon}
\affiliation{Department of Physics, Department of Information Display, and Research Institute for Basic Sciences, Kyung Hee University, Seoul, 02447, Korea}

\author{Fernando de Juan}
\affiliation{Donostia International Physics Center, P. Manuel de Lardizabal 4, 20018 Donostia-San Sebastian, Spain}
\affiliation{IKERBASQUE, Basque Foundation for Science, Maria Diaz de Haro 3, 48013 Bilbao, Spain}

\author{Zhendong Chi}
\author{F\`elix Casanova}
\affiliation{CIC nanoGUNE, 20018, Donostia-San Sebasti\'an, Basque Country, Spain}
\date{ \today }
\author{Tony Low}\email{tlow@umn.edu}
\affiliation{Department of Electrical and Computer Engineering, University of Minnesota, Minneapolis, Minnesota 55455, USA}
\affiliation{Department of Physics, University of Minnesota, Minneapolis, Minnesota 55455, USA}

\begin{abstract}
We investigate the twist angle dependence of spin-orbit coupling (SOC) proximity effects and charge-to-spin conversion (CSC) in graphene/$\textrm{WSe}_2$ heterostructures from first principles. The CSC is shown to strongly depend on the twist angle, with both the spin Hall and standard Rashba-Edelstein efficiencies optimized at or near $\ang{30}$ twisting. Symmetry breaking due to twisting also gives rise to an unconventional Rashba-Edelstein effect, with electrically generated non-equilibrium spin densities possessing spins collinear to the applied electric field. We further discuss how the carrier doping concentration and band broadening control the crossover between the Fermi-sea and -surface spin response, which reconciles the seemingly disparate experimental observations of different CSC phenomena. 
\end{abstract}

\maketitle

Graphene is an attractive channel material for spintronics owing to its long room-temperature spin diffusion length~\cite{gra-spin1,gra-spin2,gra-spin3,gra-spin4} and high carrier mobility~\cite{gra1,gra2}. However, its applicability is also strongly limited by its weak intrinsic spin-orbit coupling (SOC), which impacts the generation and manipulation of spin currents~\cite{gra-spin3}.
Recent studies have established that proximity effects can significantly enhance SOC in graphene overlaid on transition metal dichalcogenides (TMDC)~\cite{ChemSocRev.47.3359.2018,review1,RevModPhys.92.021003,proximity1, proximity2}. 
Here, the Dirac states in graphene inherit distinctive spin-textures from the proximity-induced valley-Zeeman and Rashba SOC, which gives rise to interesting phenomena such as weak anti-localization~\cite{Wang2015,Garcia2017,Yang_2016,PhysRevB.97.075434,PhysRevLett.120.106802}, giant spin lifetime anisotropy~\cite{giantspinlifetime1, giantspinlifetime2,Benitez2018} and SOC-induced spin precession~\cite{PhysRevLett.127.047202}. Transport measurements have unambiguously demonstrated efficient charge-to-spin conversion (CSC) in proximitized graphene, which can be attributed to either spin Hall or Rashba-Edelstein effects (SHE and REE, respectively)~\cite{CSC1,CSC2,Herling2020, CSC4,Li2020}. 
To date, a fundamental understanding of the controlling physics responsible for the crossover between these two regimes is still lacking.

Spin signals produced by SHE and REE can be differentiated through the direction of their generated spin polarization, which are enforced by symmetry to be mutually orthogonal to each other and transverse to the applied electric field direction~\cite{CSC4}. 
Twisted graphene/TMDC heterostructures are macroscopic chiral objects, mediated by the quantum interlayer coupling between the layers. The chirality implies that all mirror symmetries are broken, thus lifting the mutual orthogonality constraint on the allowed spin components of SHE and REE. 
Recent theoretical studies have suggested that SOC proximity effects are sensitive to the twist angle between graphene and TMDC~\cite{PhysRevB.99.075438,PhysRevB.100.085412,PhysRevB.104.195156,Pezo_2021}.
However, little is known about the impact of twist angle on the interplay of SHE and REE on CSC, in conjunction to the allowed new spin current components in this low symmetry chiral configuration.


In this letter, we address the twist angle dependence of SOC proximity effects and its connection to CSC in graphene/$\textrm{WSe}_2$ heterostructures by means of first principles calculations. 
We discovered the existence of an unconventional REE (UREE), where the spin density polarization is collinear with the applied electric field direction. 
Our results indicate that CSC is generally sensitive to the twist angle, with SHE and REE efficiencies being maximized around the 30$^{\circ}$ twist angle. 
However, it is disorder that dominantly controls the crossover between SHE and REE. 
This result is rooted in the Fermi sea (interband) and surface (intraband) characteristics of SHE and REE, respectively. Our results showed that REE dominates over SHE in the clean limit, which might help reconcile the different CSC mechanism in reported experiments.

\begin{figure}[t]
\includegraphics[width=0.95\columnwidth]{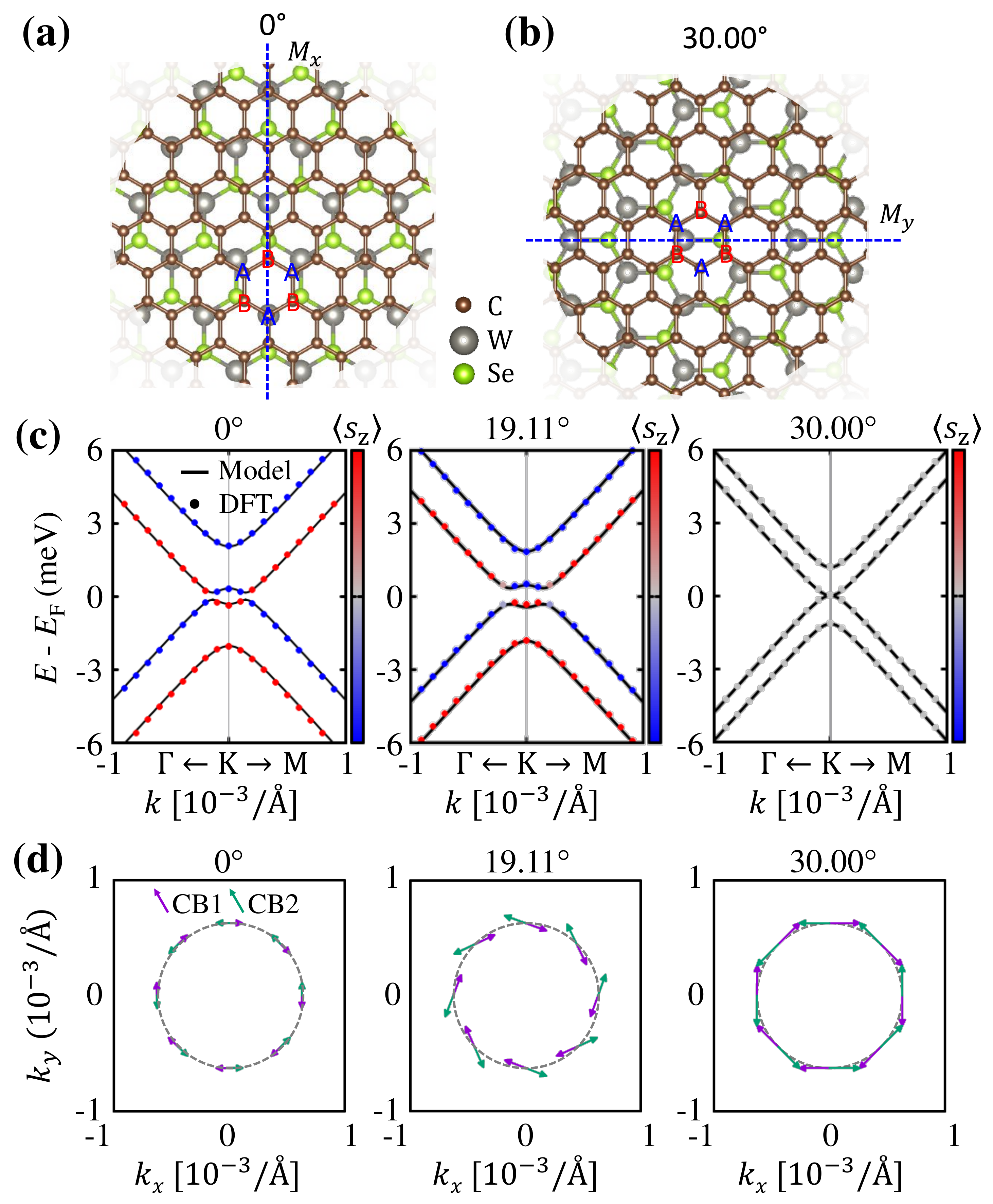}
\caption{Top views of graphene and WSe$_2$ heterostructure with (a) $0^{\circ}$ and (b) 30$^{\circ}$ twisted angles. In (a) and (b), blue dashed line indicate mirror planes and "A" and "B" highlight sublattice sites of graphene.
(c) Band structures and (d) in-plane spin distributions of heterostructures with $0^{\circ}$,  $19.11^{\circ}$ and $30^{\circ}$ twist angles.
In (c), the colored dots indicate energy eigenvalues of DFT, which were fitted by the model defined as the Eq.~(\ref{eq1}), plotted with black solid lines. The color indicates expectation value of out-of-plane spin components.
In (d), the purple and cyan arrows indicate the (CB1) lowest and (CB2) second lowest conduction bands.
}\label{fig1}
\end{figure}
We performed first-principles calculations based on the density functional theory (DFT)~\cite{Kohn1965,Kresse1996} for a total of 8 twisted graphene/$\textrm{WSe}_2$ heterostructures constructed using the coincidence lattice method~\cite{Wang2015_JPCC,PhysRevB.71.235415}.
Our lattice alignment convention is shown in Figs.~\ref{fig1}(a) and (b), where the $\theta = 0^{\circ}$ ($\theta = 30^{\circ}$) twisted heterostructure is such that the zigzag direction of graphene layer is aligned with the zigzag (armchair) direction of the WSe$_2$ lattice. It is noteworthy that the unavoidable strain originating from the artificial commensurate supercell structures can alter the relative band alignment between the graphene and the TMDC~\cite{PhysRevB.98.155309} and the SOC strength imprinted on graphene~\cite{Wang2015_JPCC,PhysRevB.104.195156}. Therefore, we carefully chose the size of the supercell structures to allow only for strain value of less than 2\% for all structures. For more detailed description of DFT calculations~\cite{Kohn1965,Kresse1996}, see Supplementary Information (SI)~\cite{Snote}.
In all twisted angles, the Dirac cones of graphene lie within the WSe$_2$ band gap, which guarantees that charge and spin transport in twisted graphene/$\textrm{WSe}_2$ is Dirac-like at low dopings~\cite{proximity2, REEPRL}. (See also Fig.S2) 

We now analyse the proximity-induced spin texture by performing fully spin-orbit coupled calculations. Figure~\ref{fig1}(c) shows the electronic structures of $0^{\circ}$, $19.11^{\circ}$ and $30^{\circ}$ twisted graphene/$\textrm{WSe}_2$ heterostructures. We observe well defined out-of-plane spin-polarized sub-bands and clearly inverted band structures for all $\theta \neq 30^{\circ}$ twistings~\cite{Snote}. The in-plane spin textures for the three heterostructures are shown in Fig.~\ref{fig1}(d). It exhibits a chiral Rashba-like spin-momentum locking superposed with an additional out-of-plane spin texture for all $\theta \neq 30^{\circ}$, implying the existence of valley-Zeeman SOC~\cite{Snote}. 
Here, the spin states gradually tilt toward the in-plane direction with increasing $\theta$, where $\langle s_z \rangle$ becomes completely quenched at 30$^{\circ}$ for all sub-bands. 
The removal of mirror plane symmetries at most twist angles also imbued the spin texture with additional features. Unlike ordinary Rashba spin splitted two-dimensional electron gas, herein, the in-plane spins are not perfectly orthogonal to the electron’s momentum~\cite{PhysRevB.99.075438,PhysRevB.100.085412}.


\begin{figure}[t]
\includegraphics[width=0.95\columnwidth]{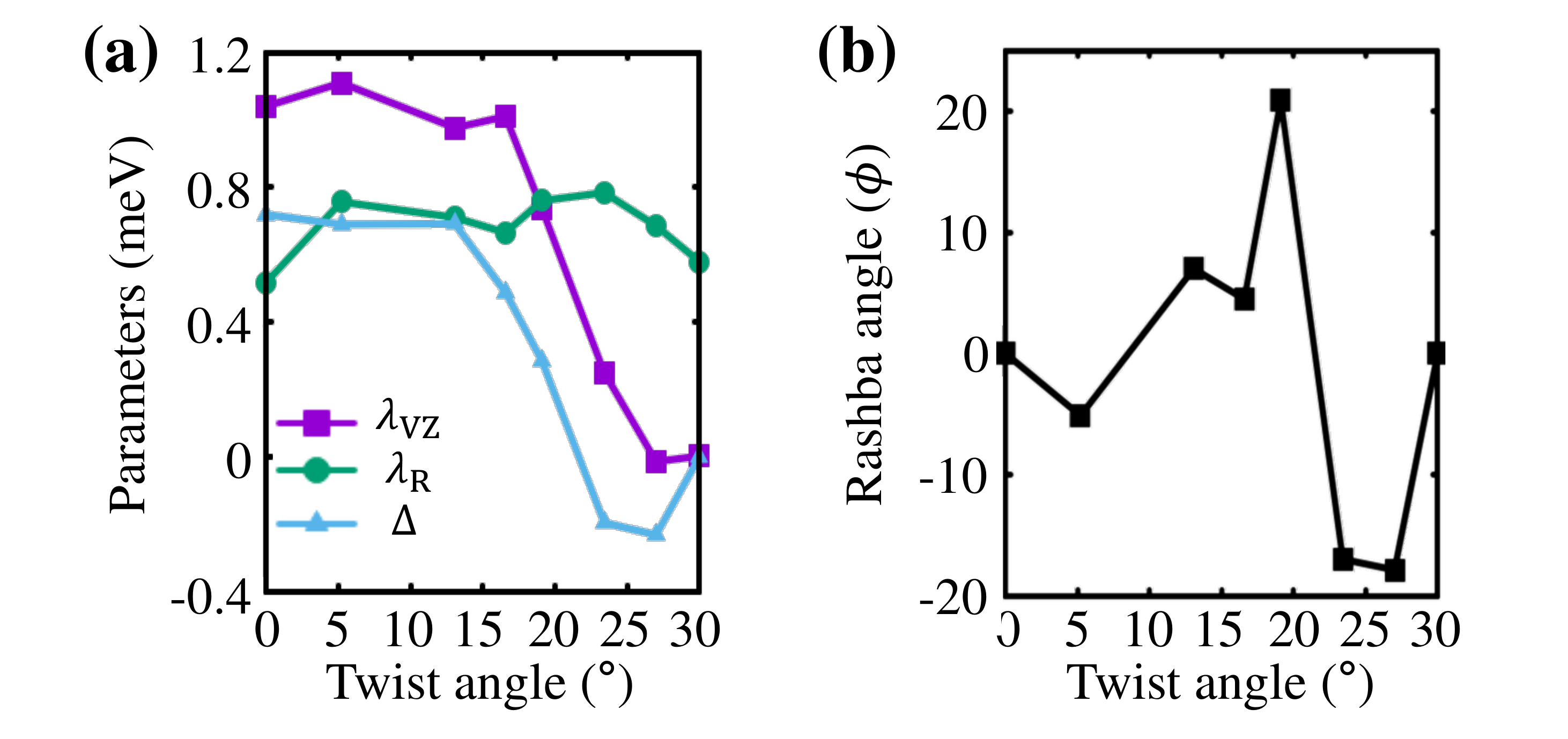}
\caption{The twist angle evolution of (a) parameters of model Hamiltonian as described in the Eq.~(\ref{eq1}) and (b) corresponding Rashba angle.
In (a), purple, cyan, and skyblue lines indicate valley-Zeeman ($\lambda_{\mathrm{VZ}}$), Rashba SOC ($\lambda_{\mathrm{R}}$) and staggerd potential ($\Delta$), respectively.
}\label{fig2}
\end{figure}

To obtain the twist angle evolution of the proximity-induced SOC parameters, the band structure and spin texture were fitted to the continuum Hamiltonian~\cite{tightbinding}

\begin{eqnarray}
& H(\kappa \textbf{K} + \textbf{k}) =  \hbar v_F (\kappa \sigma_x k_x + \sigma_y k_y) + \Delta \sigma_z + \nonumber\\
 & + \lambda_R e^{-i s_z \phi/2}(\kappa \sigma_x s_y - \sigma_y s_x)e^{i s_z \phi/2} + \nonumber\\
 & + (\lambda_{\rm{VZ}}\sigma_0 +\lambda_{\rm{KM}}\sigma_z)\kappa s_z ,
\label{eq1}
\end{eqnarray}
where $v_F$ is the Fermi velocity in graphene, $\Delta$ is the sublattice asymmetry and $\kappa = \pm 1$ is the valley index (for valley $\pm \textbf{K}$). 
The remaining parameters $\lambda_{\rm{R}}$, $\lambda_{\rm{VZ}}$, and $\lambda_{\rm{KM}}$ collectively account for the proximity-induced SOC and describe, respectively, the Rashba, valley-Zeeman, and Kane-Mele SOC terms.
The last two terms also can be understood as antisymmetric and symmetric part of intrinsic SOC ($\lambda_I$) of two sublattices (A and B) of graphene, or  $\lambda_{\rm{VZ}}=(\lambda_I^{A}-\lambda_I^{B})/2$, and $\lambda_{\rm{KM}}=(\lambda_I^{A}+\lambda_I^{B})/2$, respectively.
Finally, the $\sigma_i$ ($s_i$) matrices, with $i = 0, x, y, z$, operate on the orbital (spin) space and $\phi$ is the Rashba angle parameter accounting for the non-orthogonal spin-momentum locking~\cite{PhysRevB.99.075438,PhysRevB.100.085412,PhysRevB.104.195156,Pezo_2021}.
Note that, as reported in earlier works~\cite{PhysRevB.99.075438,PhysRevB.100.085412,PhysRevB.104.195156}, we also found $\lambda_{\rm{KM}}$ to be negligibly small and does not affect the CSC.

Figure~\ref{fig1}(c) shows the excellent agreement between the first principles (symbols) and continuum model (solid) bands. The twist angle evolution of valley-Zeeman $\lambda_{\rm{VZ}}$, $\lambda_{\rm{R}}$ and $\Delta$ are summarized in Fig.~\ref{fig2}(a). 
We observe that $\lambda_{\rm{VZ}}$ is larger than $\lambda_{\rm{R}}$ at small twisting angles, and both $\lambda_{\rm{VZ}}$ and $\Delta$ vanish at the 30$^{\circ}$ twist angle, as required by symmetries.
This behavior is related to the existence of mirror planes in certain twisted graphene/$\textrm{WSe}_2$ systems and to the sublattice symmetry of graphene: Due to the $C_{\rm{6}}$ and $C_{\rm{3}}$ rotation symmetries of graphene and WSe$_2$, respectively, the $0^{\circ}$ ($30^{\circ}$)-twisted graphene/$\textrm{WSe}_2$ heterostructures possess a $\mathbf{M_x}$ ($\mathbf{M_y}$) mirror, as shown in Figs.~\ref{fig1}(a) and (b), where we also highlight sublattice sites of graphene as ``A'' and ``B''. At $\theta=0^{\circ}$ ($\theta=30^{\circ}$) twisting, sublattices ``A'' and ``B'' are left (are not left) invariant under a $\mathbf{M_x}$ ($\mathbf{M_y}$) mirror operation. 
Hence, both $\lambda_{\rm{VZ}}$ and $\Delta$ vanish at $\theta=30^{\circ}$.
In contrast, $\lambda_{\rm{R}}$ exhibits only a slight twist angle modulation, with a magnitude that is approximately the same for both $0^{\circ}$ and $30^{\circ}$.
The twist angle evolution of the Rashba-angle, $\phi$, is summarized in Fig.~\ref{fig2}(b). The existence of $\mathbf{M_x}$ ($\mathbf{M_y}$) mirror planes constraints $\phi$ to be zero at $0^{\circ}$ ($30^{\circ}$) twistings~\cite{PhysRevB.99.075438,PhysRevB.100.085412}. Symmetry breaking at all other twist angles enables the existence of finite $\phi$.
In addition, we find that the magnitude and sign of $\phi$ vary rapidly with the twist angle, resulting in the angle-dependent UREE, which will be discussed below.

\begin{figure}[t]
\includegraphics[width=\columnwidth]{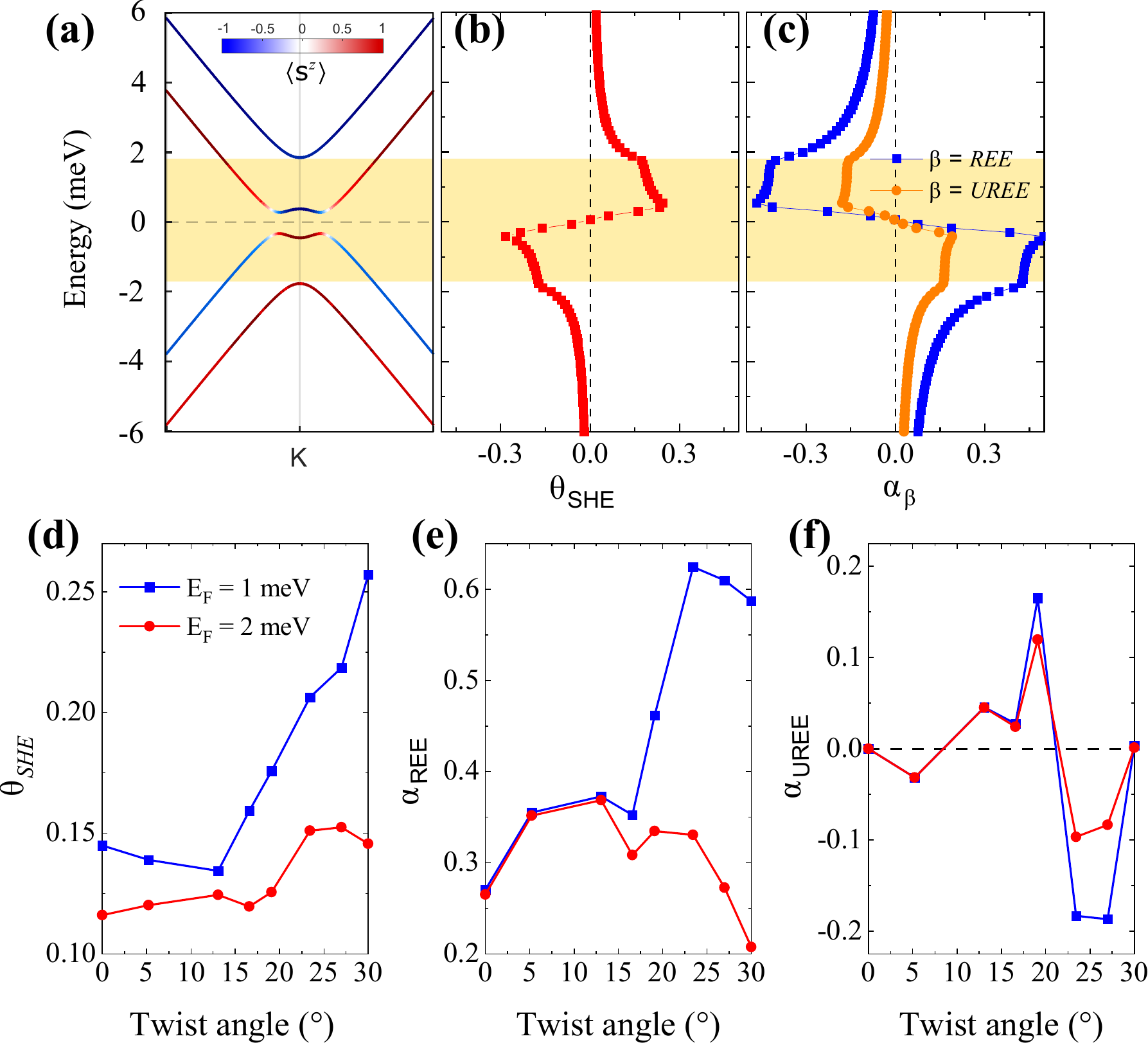}
\caption{(a) Spin-resolved band structure of a graphene/$\textrm{WSe}_2$ heterostructure with $\theta = 19.11^{\circ}$. Colors refer to the $\langle s^z \rangle$ magnitude. The associated spin Hall, Rashba-Edelstein and unconventional Rashba-Edelstein efficiencies, $\theta_{\textrm{SHE}}$, $\alpha_{\textrm{REE}}$ and $\alpha_{\textrm{UREE}}$ respectively, are shown in panels (b) and (c). The horizontal yellow stripes highlight the gap between conduction $\langle s^z \rangle = -1$ and valence $\langle s^z \rangle = +1$ sub-bands. The twist angle evolution of CSC efficiencies are displayed in panels (d), (e) and (f) at two distinct doping levels.}
\label{fig3}
\end{figure}

Next, we study how the twist angle modulation of the proximity-induced SOC affects CSC. 
As previously discussed, CSC in twisted graphene/$\textrm{WSe}_2$ is solely due to the proximitized Dirac cones at low doping levels, hence it vanishes when $E_{\rm{F}}$ $\gg$ $\lambda_{\rm{VZ}}$ and $\lambda_{\rm{R}}$, where $E_{\rm{F}}$ is the Fermi energy. 
To account for that, we write down a fully periodic tight-binding Hamiltonian to investigate the twist angle evolution of SHE and REE~\cite{Snote}.
The electronic response to external electric fields is treated within the linear response theory. Here, we utilize the Kubo formula fashioned after Smrcka-Streda~\cite{kubo1, kubo2, kubo3, kubo4}:
\begin{eqnarray}
     & \delta O_{\alpha \beta}^{\gamma} = \displaystyle \int \frac{d\textbf{k}}{(2\pi)^2} [\delta O_{\alpha \beta}^{\gamma, I}(\textbf{k}) + \delta O_{\alpha \beta}^{\gamma, II}(\textbf{k})],
     \label{eq2}
\end{eqnarray}
with integrands 
\begin{subequations}
\begin{equation}
\resizebox{\hsize}{!}{$\delta O_{\alpha \beta}^{{\gamma}, I}(\textrm{k}) = \displaystyle -\frac{e \hbar}{\pi} {\Gamma}^2 \sum_{nm} \frac{\operatorname{Re}
[\langle n\textbf{k}|\hat{O}_{\alpha}^{\gamma}|m\textbf{k} \rangle \langle m\textbf{k}| \hat{v}_{\beta}| n \textbf{k}\rangle]}{[({\epsilon}_{F} - {\epsilon}_{n{\textbf{k}}})^2+{\Gamma}^2][({\epsilon}_{F} - {\epsilon}_{m{\textbf{k}}})^2+{\Gamma}^2]}$},
\label{eq3a}
\end{equation}
\begin{equation}
\resizebox{\hsize}{!}{$\delta O_{\alpha \beta}^{\gamma, II}(\textrm{k}) = \displaystyle -2e\hbar \sum_{n,m\neq n}(f_{n\textbf{k}} - f_{m\textbf{k}}) \frac{\operatorname{Im}[\langle n\textbf{k}|\hat{O}_{\alpha}^{\gamma}|m\textbf{k} \rangle \langle m\textbf{k}| \hat{v}_{\beta}| n\textbf{k}\rangle]}{({\epsilon}_{n\textbf{k}} - {\epsilon}_{m\textbf{k}})^2  - {\Gamma}^2}$},
\label{eq3b}
\end{equation}
\end{subequations}
where $\hat{v}_{\beta}$ is the $\beta = x,y,z$ component of the velocity operator, $\hat{O}_{\alpha}^{\gamma}$ is the perturbed physical observable with spin index $\gamma = x,y,z$ and $| n\textbf{k}\rangle$ is the eigenstate associated with the band $\epsilon_{n\textbf{k}}$ of the unperturbed system. 
The Smrcka-Streda formula provides a good description in the weak disorder limit, which is assumed to only cause a constant band broadening quantified by $\Gamma$, and naturally includes both Fermi surface and Fermi sea contributions in Eqs.~(\ref{eq3a}) and (\ref{eq3b}) respectively~\cite{kubo1}. 
The charge, spin Hall and spin density responses are obtained through $\hat{O}_{\alpha}^{\gamma} \rightarrow -e\hat{v}_{\alpha}$, $\hat{O}_{\alpha}^{\gamma} \rightarrow (2/\hbar)\hat{Q}_{\alpha}^{\gamma}$, where the spin current operator is defined as $\hat{Q}_{\alpha}^{\gamma} = (1/2)\{\hat{s}^{\gamma}, \hat{v}_{\alpha}\}$ with spin operator $\hat{s}^{\gamma}$, and $\hat{O}_{\alpha}^{\gamma} \rightarrow \hat{s}^{\gamma}$, respectively. In the following, we assume a constant electric field applied along the $\hat{\textbf{x}}$ direction and define the SHE, REE and UREE efficiencies as $\theta_{\textrm{SHE}} = (2e/\hbar) \sigma_{yx}^z/\sigma_{xx}$, $\alpha_{\textrm{REE}} = (2ev_F/\hbar)\delta s^{y}/\sigma_{xx}$ and $\alpha_{\textrm{UREE}} = (2ev_F/\hbar)\delta s^{x}/\sigma_{xx}$, where 
$\sigma_{yx}^z$ and $\sigma_{xx}$ are the spin hall and charge conductivities, 
$\delta s^{y}$ and $\delta s^{x}$ are the electrically-induced spin densities,
and $v_F = 1\times 10^6$ m/s is the Fermi velocity in graphene~\cite{REEPRL}. 

\begin{figure}[t]
\includegraphics[width=\columnwidth]{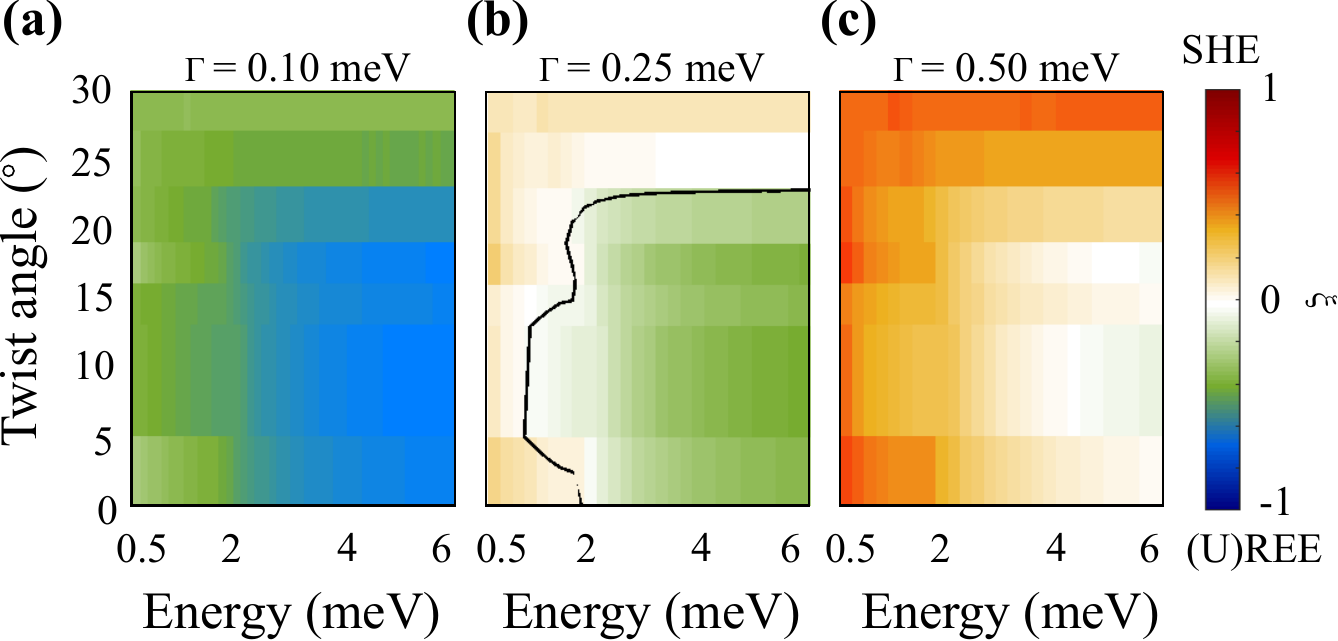}
\caption{Evolution of the CSC diagram with the band broadening: (a) $\Gamma = 0.10$ meV, (b) $\Gamma = 0.25$ meV and (c) $\Gamma = 0.50$ meV. Here, $\xi(E, \theta) = (|\theta_{\textrm{SHE}}|-|\alpha|)/(|\theta_{\textrm{SHE}}|+|\alpha|)$ is a CSC efficiency polarization, where $\alpha = \sqrt{\alpha_{\textrm{REE}}^2 + \alpha_{\textrm{UREE}}^2}$, such that stronger positive (negative) regions imply in dominant CSC through SHE (REE or UREE). The contour in panel (b) separates dominating SHE and REE/UREE scenarios where $\xi = 0$.}
\label{fig4}
\end{figure}

Figure~\ref{fig3}(a) shows the spin-resolved bands of $19.11^{\circ}$ twisted graphene/$\textrm{WSe}_2$ at the vicinity of the $K$ point. The associated energy-resolved SHE and REE efficiencies, shown in Figs.~\ref{fig3}(b) and (c) considering $\Gamma = 0.1$ meV, reveal that efficient CSC takes place within a small energy window between the valence $\langle s^z \rangle = +1$ and conduction $\langle s^z \rangle = -1$ sub-bands [shaded yellow region in Figs.~\ref{fig3}(a), (b) and (c)]. The presence of additional sub-bands at higher energies acts as to suppress the CSC due to their opposite contributions to the total spin-Berry curvature and opposite spin-momentum locking helicity.
Figure~\ref{fig3}(c) also shows a sizable UREE whose efficiency is comparable to that of SHE.
The observation of such novel spin currents, whose spin quantization axis and electric field direction are collinear, has been recently reported in the in graphene/MoTe$_2$~\cite{Safeer2019}, graphene/WTe$_2$~\cite{Camosi_2022}, and graphene/NbSe$_2$~\cite{https://doi.org/10.48550/arxiv.2205.07668} heterostructures.

We explore the twist angle evolution of the CSC efficiencies at two distinct doping levels in Figs.~\ref{fig3}(d), (e) and (f). Here, the maximum SHE efficiency for the higher doping case, $E_{\rm{F}} = 2$ meV, occurs between $23^{\circ}$ and $27^{\circ}$ twisting. The situation differs substantially at lower doping levels or $E_{\rm{F}} = 1$ meV, where our results indicate a larger SHE efficiency at $30^{\circ}$ twisting. Similar to the spin Hall case, the REE efficiency is also more sensitive to the twist angle at lower doping levels, as shown in Fig.~\ref{fig3}(e). 
A similar behavior is observed for the UREE in Fig.~\ref{fig3}(f). Remarkably, both REE and UREE become more efficient in the lowest
doping case exhibiting a maximum at $\theta \approx 23^{\circ}$ twisting. We also found that while the conventional REE efficiency remains sizable and finite at $\theta \approx 30^{\circ}$, the UREE abruptly vanishes due to symmetry constraints on the Rashba angle $\phi$ with the restoration of a mirror plane. 
The absence of $\lambda_{\rm{VZ}}$ at $30^{\circ}$ twisting suggests that the valley-Zeeman SOC is detrimental to SHE-based CSC efficiency in the clean limit. This is due to the fact that the charge conductivity $\sigma_{xx}$ increases faster than $\sigma_{yx}^z$ with $\lambda_{\rm{VZ}}$~\cite{Snote}. 

Although SHE and REE type-of-CSC are simultaneously present in graphene/$\textrm{WSe}_2$, their relative ability to produce spins might vary with doping, twist angle and disorder. 
To facilitate this study we define a CSC efficiency polarization as $\xi(E, \theta) = (|\theta_{\textrm{SHE}}|-|\alpha|)/(|\theta_{\textrm{SHE}}|+|\alpha|)$ where $\alpha = \sqrt{\alpha_{\textrm{REE}}^2 + \alpha_{\textrm{UREE}}^2}$, and tracked its evolution with the band broadening $\Gamma$ in Fig.~\ref{fig4}. For small band broadening ($\Gamma = 0.1$ meV), $\xi$ is mostly negative for all energies and twist angles, i.e., CSC is dominated by the REEs [See Fig.~\ref{fig4}(a)]. At large spectral broadening ($\Gamma = 0.50$ meV), $\xi$ turned positive throughout, as shown in  Fig.~\ref{fig4}(c), indicating that SHE dominates the CSC. We identify an intermediate crossover broadening ($\Gamma = 0.25$ meV), the dominating mechanism for CSC strongly depends on the twist angle and doping levels, being mostly due to SHE close to $30^{\circ}$ twisting and to REE at higher doping levels for twisting angles in the interval $5^{\circ}-25^{\circ}$, as shown in Fig.~\ref{fig4}(b). 

The strong $\Gamma$ dependence of $\xi$ originates from the contrasting Fermi sea and Fermi surface nature of SHE and REEs, respectively. Here, $\delta s^{x, y}$ and $\sigma_{xx}$ are both Fermi surface effects, such that their dependences on $\Gamma$ compensate and give rise to an $\alpha$ that is approximately independent of $\Gamma$. On the contrary, the SHE is approximately independent of $\Gamma$ because it originates from the Fermi sea. Hence, $\sigma_{yx}^{z}$ does not compensate the $\Gamma$ dependence of $\sigma_{xx}$. 
These findings shed light on the disparate dominant CSC mechanism reported across different proximitized graphene Hall bar devices.
For examples, in the recent experiments, REE is dominant over SHE in WS$_{\rm{2}}$/graphene/hBN/SiO$_{\rm{2}}$/Si Hall bar devices~\cite{CSC1}, but the opposite is true in MoS$_{\rm{2}}$ or WSe$_{\rm{2}}$/graphene/SiO$_{\rm{2}}$/Si devices~\cite{CSC2, Herling2020}, which is consistent with the physical picture we presented, that REE dominates in the clean limit. However, more robust validation would require more experimental studies.

In summary, we have studied proximity effects and its relation to the charge-to-spin (CSC) conversion in twisted graphene/WSe$_2$ heterostructures from first principles. 
We have analysed in detail how the SHE and REE efficiencies are affected by the twist angle and found that optimal CSC of in-plane and out-of-plane spins occurs for structures with around $30^{\circ}$ twisting.
In addition, our results revealed that lack of mirror symmetry for $0^{\circ}<\theta<30^{\circ}$ twisted structures leads to non-orthogonal Rashba spin texture, resulting in spin accumulation longitudinal to the applied electric field. We also addressed the question of dominant CSC mechanism by studying how SHE and REE are affected by spectral broadening, carrier doping levels and twist angle. Our work highlights the new physics of CSC in graphene/TMDC heterostructures brought about by twistronics.

\textit{Acknowledgments} 
S. L. is primarily supported by Basic Science Research Program through the National Research Foundation of Korea funded by the Ministry of Education (NRF-2021R1A6A3A14038837).
S. L. and T. L. are partially supported by NSF DMREF-1921629.
D. S. and T. L. are partially supported by the Valleytronics Intel Science Technology Center, and SMART, one of the seven centers of nCORE, a Semiconductor Research Corporation program, sponsored by National Institute of Standards and Technology (NIST).
Y.K. acknowledges financial support from the Korean government through the National Research Foundation of Korea (NRF-2022M3F3A2A01073562).
Z.C. and F.C. acknowledge funding by the Valleytronics Intel Science Technology Center, the Spanish MICINN (Project RTI2018-094861-B-I00 and Maria de Maeztu Units of Excellence Programme CEX2020-001038-M) and by the Regional Council of Gipuzkoa (Project ORBILOGICS)
F.J. acknowledges funding from the Spanish MCI/AEI/FEDER through grant PGC2018-101988-B-C21 and from Diputacion de Gipuzkoa through grant Gipuzkoa NEXT 2021-100-000070-01.
We acknowledge the MSI in the University of Minnesota for providing the computational resources, and useful discussions with Raseong Kim and Ian Young from Intel Corporation.


%
%
%
%

\end{document}



\title{{\normalsize{Supplementary Information}} \\
Charge-to-spin conversion in twisted graphene/$\textrm{WSe}_2$ heterostructures}

\author{Seungjun Lee}\thanks{These authors contributed equally to this work.}
\author{D. J. P. de Sousa}\thanks{These authors contributed equally to this work.}
\affiliation{Department of Electrical and Computer Engineering, University of Minnesota, Minneapolis, Minnesota 55455, USA}

\author{Young-Kyun Kwon}
\affiliation{Department of Physics, Department of Information Display, and Research Institute for Basic Sciences, Kyung Hee University, Seoul, 02447, Korea}

\author{Fernando de Juan}
\affiliation{Donostia International Physics Center, P. Manuel de Lardizabal 4, 20018 Donostia-San Sebastian, Spain}
\affiliation{IKERBASQUE, Basque Foundation for Science, Maria Diaz de Haro 3, 48013 Bilbao, Spain}

\author{Zhendong Chi}
\author{F\`elix Casanova}
\affiliation{CIC nanoGUNE, 20018, Donostia-San Sebasti\'an, Basque Country, Spain}
\date{ \today }

\author{Tony Low}\email{tlow@umn.edu}
\affiliation{Department of Electrical and Computer Engineering, University of Minnesota, Minneapolis, Minnesota 55455, USA}
\affiliation{Department of Physics, University of Minnesota, Minneapolis, Minnesota 55455, USA}

\date{\today}
\maketitle

\subsection*{Note S1: Computational details}
\label{note1}

We performed first-principles calculations based on density functional theory (DFT)~\cite{Kohn1965} as implemented in Vienna \textit{ab initio} simulation package (VASP)~\cite{Kresse1996}. 
The exchange-correlation (XC) functional was treated within the generalized gradient approximation of Perdew-Burke-Ernzerhof (PBE)~\cite{Perdew1996} with noncollinear spin polarization.~\cite{PhysRevB.93.224425}
The electronic wavefunctions were expanded by planewave basis with kinetic energy cutoff of 400~eV. We employed the projector-augmented wave pseudopotentials{~\cite{{Blochl1994},{Kresse1999}}} to describe the valence electrons, and Grimme-D2 Van der Waals correction~\cite{grimme-d2} to describe long range interaction between graphene and WSe$_2$.
The sufficiently large vacuum region was included to mimic 2D layered or slab structure in periodic cells.

To construct supercell configurations with various twist angle, we employed accidental angular
commensuration method.~\cite{Wang2015_JPCC,PhysRevB.71.235415}
In the hexagonal two-dimensional materials, any supercell structrue $(m, n)$ can be defined by two lattice vectors $\mathbf{a_{\rm{(m,n)}}}=m \mathbf{a} + n \mathbf{b}$ and $\mathbf{b_{\rm{(m,n)}}}=-m \mathbf{a} + (m+n) \mathbf{b}$ with a corresponding skewed angle $\theta_{\rm{(m,n)}}=tan^{-1}(\sqrt{3}m/(2n+m))$.
Based on the equilibrium lattice constants of graphene ($a^{\rm{Gra}}=2.46~\rm{\AA}$) and WSe$_2$ ($a^{\rm{WSe_2}}=3.32~\rm{\AA}$), we constructed supercells of graphene $(m,n)$ and WSe$_2$ $(m',n')$, and stacked them to make heterostructures while keeping their lattice mismatch below 2\%. The twist angle is obtained by $\theta=|\theta^{\rm{Gra}}_{\rm{(m,n)}}-\theta^{\rm{WSe_2}}_{\rm{(m',n')}}|$, and the lattice constants of heterostructure was used to be the same value as $a^{\rm{Gra}}_{\rm{(m,n)}}$.
Within the given lattice constants, the atomic positions were fully relaxed through DFT calculation.
The detailed structural parameters and corresponding $k$-mesh gird ($N_k \times N_k$) of the heterostructures are summarized in Table~\ref{table1}.

The band ($n$) and momentum ($\mathbf{k}$) resolved spin angular momentum distributions are calculated by the expectation value of the spin angular momentum operator, that is,  $\sum_{\alpha}\mathinner {\protect \langle {\psi^\alpha _{n,\protect \mathbf {k}} |\protect \bm{\sigma} | \psi^\alpha _{n,\protect \mathbf {k}}}\protect \rangle }$, where $\protect \bm {\sigma }$ is the Pauli spin matrix vector, and $\psi^\alpha_{n,\mathbf{k}}$ is an wavefunction projected on atom site $\alpha$.

\subsection*{Note S2: Quantum transport calculation details}
\label{note2}

The quantum transport calculations are based on a periodic tight-binding model for the proximitized graphene Dirac states. The Hamiltonian reads~\cite{tightbinding}
\begin{eqnarray}
     & H = \displaystyle \sum_{\langle ij\rangle, \sigma} \hat{c}_{i\sigma}^{\dagger} t \hat{c}_{j\sigma} + \sum_{i, \sigma} \hat{c}_{i\sigma}^{\dagger} \Delta \xi_{c_{i}} \hat{c}_{i\sigma} + \nonumber \\
     & \displaystyle\frac{2i\lambda_{\rm{R}}}{3}\sum_{\langle ij\rangle, \sigma \sigma'} \hat{c}_{i\sigma}^{\dagger} [e^{-i\hat{s}_z \phi/2}(\hat{\textbf{s}}\times \hat{\textbf{d}}_{ij})\cdot \hat{\textbf{z}} e^{i\hat{s}_z \phi/2}]_{\sigma \sigma'} \hat{c}_{j\sigma} \nonumber \\
     & \displaystyle + \frac{i}{3}\sum_{\langle\langle ij\rangle\rangle, \sigma \sigma'} \hat{c}_{i\sigma}^{\dagger} \left[\frac{\lambda_{I}^{c_i}}{\sqrt{3}}\nu_{ij}s_z + 2\lambda_{\rm{PIA}}^{c_i}(\hat{\textbf{s}}\times \hat{\textbf{D}}_{ij})\cdot \hat{\textbf{z}}\right]_{\sigma \sigma'} \hat{c}_{j\sigma},
     \label{eq1}
\end{eqnarray}
where $\hat{c}_{i\sigma}^{\dagger}$ ($\hat{c}_{i\sigma}$) creates (annihilates) a $p_z$ electron with spin $\sigma$ at site $i$, $t$ is the nearest neighbor hopping parameter and $\Delta$ is the sublattice asymmetry with $\xi_{A(B)}=\pm 1$. The remaining parameters $\lambda_{\rm{R}}$, $\lambda_{\rm{I}}^{\rm{A,B}}$ and $\lambda_{\rm{PIA}}^{\rm{A,B}}$ collectively account for the proximity-induced SOC and describe, respectively, the Rashba SOC, the sublattice resolved intrinsic SOC and pseudospin inversion asymmetry induced terms.
In addition $\nu_{ij} = +1 (-1)$ describes the relative phase acquired by an electron as it travels from sites $i$ and $j$ in a clockwise (counterclockwise) sense and $\hat{d}_{ij}$ and $\hat{D}_{ij}$ are, respectively, the nearest neighbor and next-nearest neighbor unit vectors connecting sites $i$ and $j$. Finally, $\hat{s}$ is the vector of spin Pauli matrices and $\phi$ is the Rashba angle parameter accounting for the non-orthogonal spin-momentum locking~\cite{PhysRevB.99.075438,PhysRevB.100.085412,PhysRevB.104.195156,Pezo_2021}. The above Hamiltonian can be Fourier transformed to momentum space, from where eigenstates and energy bands relevant for the linear response formula are obtained.

Figure~\ref{figs6} displays the energy dependence of the spin Hall conductivity and Rashba-Edelstein efficiency (up to a factor of $v_F$) for all twist angles considered in this work. From Fig.~\ref{figs6}(a) it is clear that the spin Hall conductivity is maximized at $30^{\circ}$ twisting at all energies. This implies that the Rashba-like spin texture in this situation is beneficial to the spin Hall effect. The Rashba-Edelstein efficiency in Fig.~\ref{figs6}(b) shows that the maximum efficiency varies with energy, where maximum efficiency occurs at the vicinity of $30^{\circ}$ twisting at low doping levels. 

Figure~\ref{figs7} is a study on the relevance of the proximity-induced pseudospin inversion asymmetry on the charge-to-spin conversion. We considered two situations, where we artificially set $\lambda_{\rm{PIA}}^{A,B}$ to zero or maintained its value to that obtained by fitting the first-principles ground states. Panels (a) and (b) show that the same behavior is expected regardless of the $\lambda_{\rm{PIA}}^{A,B}$ we assume. The same conclusion is valid for all twist angles we considered (two sample twist angle cases are shown). Hence, the pseudospin inversion asymmetry SOC due to procimity-effects is not relevant to charge-to-spin conversion in Gra/TMDC heterostructures. 

Figure~\ref{figs8} is a sample calculation showing that the valley Zeeman SOC parameter is detrimental to the charge-to-spin conversion efficiencies. In these calculations, we artificially increase the absolute values of the parameters $\lambda_{I}^{A,B}$ while maintaining the remaining parameters fixed at those corresponding to the $0^{\circ}$ twisted structure. As is seen, the maximum efficiency decays and is shifted to higher energies at larger $\lambda_{VZ}$ for both spin Hall and Rashba-Edelstein effects. The energy shift is due to the band structure modification associated with $\lambda_{I}^{A,B}$, while the suppression is mainly due to the larger longitudinal conductivity $\sigma_{xx}$ at larger $\lambda_{VZ}$. It is worth emphasizing that these results are valid in the weak disorder limit, where the constant $\Gamma$ approximation provides a good description. In the strong disorder limit, the relation between efficiency and valley-Zeeman SOC is qualitatively distinct.


\clearpage
\newpage
\begin{table}
\caption {Structural parameters of the twisted graphene/WSe$_2$ heterostructure. The lattice constants of heterostructure was used to be the same value as $a^{\rm{Gra}}_{\rm{(m,n)}}$, and the residual strain was only applied to the WSe$_2$ $(m',n')$.
\label{table1}}
\begin{ruledtabular}
\begin{tabular}{ c c c c c c c c c c c } 
   $\theta$ ($^{\circ}$) & Strain ($\%$) & $N_k$ & $a^{\rm{Gra}}_{\rm{(m,n)}}$ & $m$ & $n$ & $\theta^{\rm{Gra}}_{\rm{(m,n)}}$ & $a^{\rm{WSe_2}}_{\rm{(m',n')}}$ & $m'$ & $n'$ & $\theta^{\rm{WSe_2}}_{\rm{(m',n')}}$  \\
    \hline
0.00  &  1.20   & 6 & 9.84   & 4   & 0   & 0.00    & 9.96     &  3  &  0  &   0.00 \\
5.21  &  -0.98  & 3 &  15.36 &   5 &   2 &   16.10 &    15.21 &   4 &   1 &    10.89 \\
13.07 &   0.29  & 2 &  20.14 &   7 &   2 &   12.22 &    20.19 &   4 &   3 &    25.29 \\
16.58 &   -0.98 & 2 &   23.47&    9&    1&    5.21 &    23.24 &   5 &   3 &    21.79 \\
19.11 & -1.98 & 4 & 13.02  & 4 &  2 &  19.11  & 13.28  & 4 &  0 &  0 \\
23.41 & 2.00  & 4 & 14.76 &  6 &  0 &  0 &  14.47 &  3 &  2 &  23.41 \\
27.00 &   -0.98 & 2 &   23.47&    6&    5&    27.00&     23.24&    7&    0&     0.00 \\
30.00 &   0.18  & 2 &  17.22 &   7 &   0 &   0.00  &   17.25  &  3  &  3  &   30.00 \\
\end{tabular}
\end{ruledtabular}
\end{table}

\clearpage
\newpage
\begin{figure}[p]
\includegraphics[width=1.0\columnwidth]{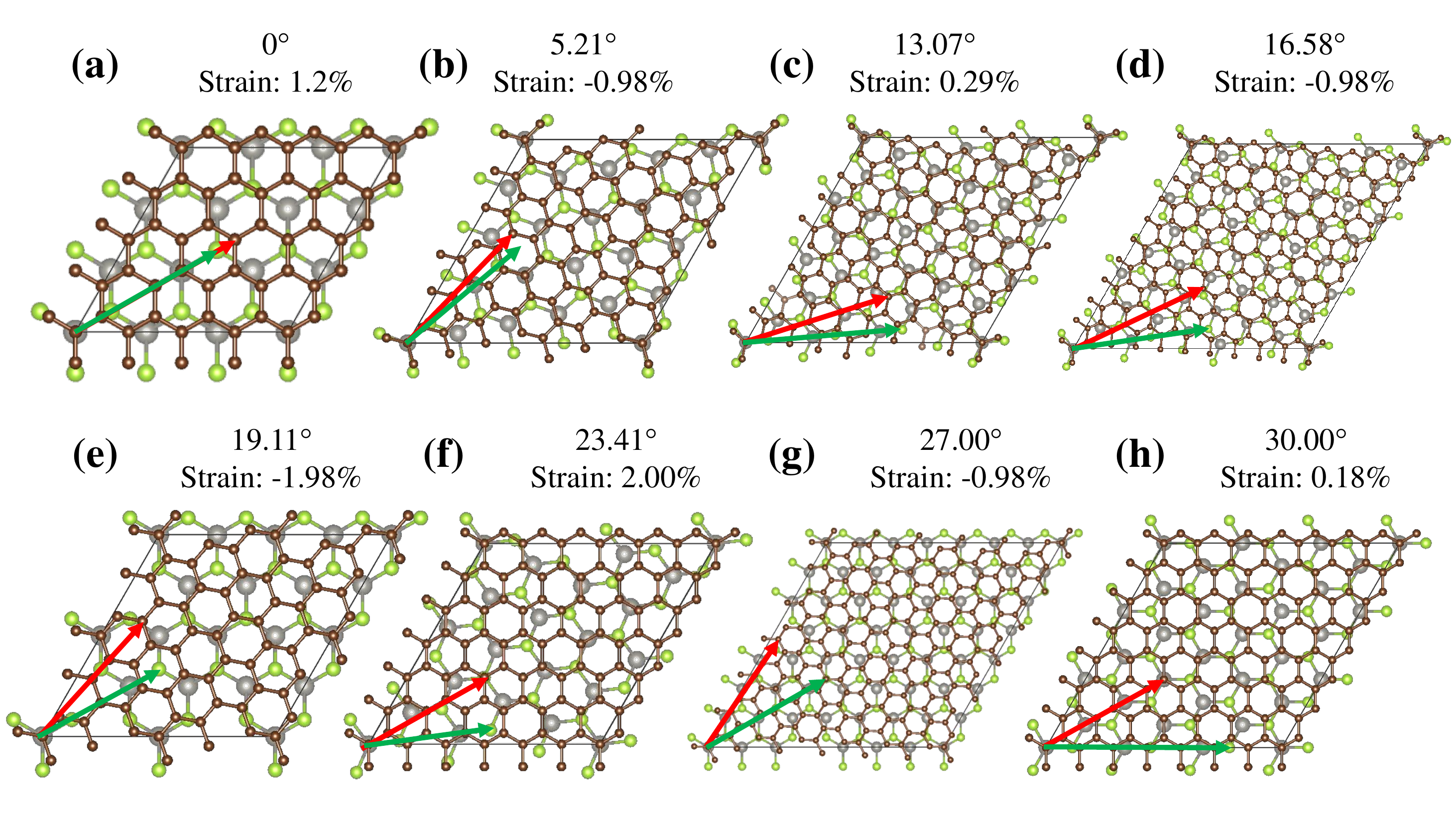}
\caption{(a--h) Crystal structure of twisted graphene/WSe$_{\rm{2}}$ heterostructures with a total of 8 twist angles. Brown, grey, and green spheres indicate carbon, tungsten, and selenium atoms, respectively.
In each figure, red and green arrows show the armchair directions of graphene and  WSe$_2$.}
\label{figs1}
\end{figure}
\clearpage
\newpage
\begin{figure}[p]
\includegraphics[width=1.0\columnwidth]{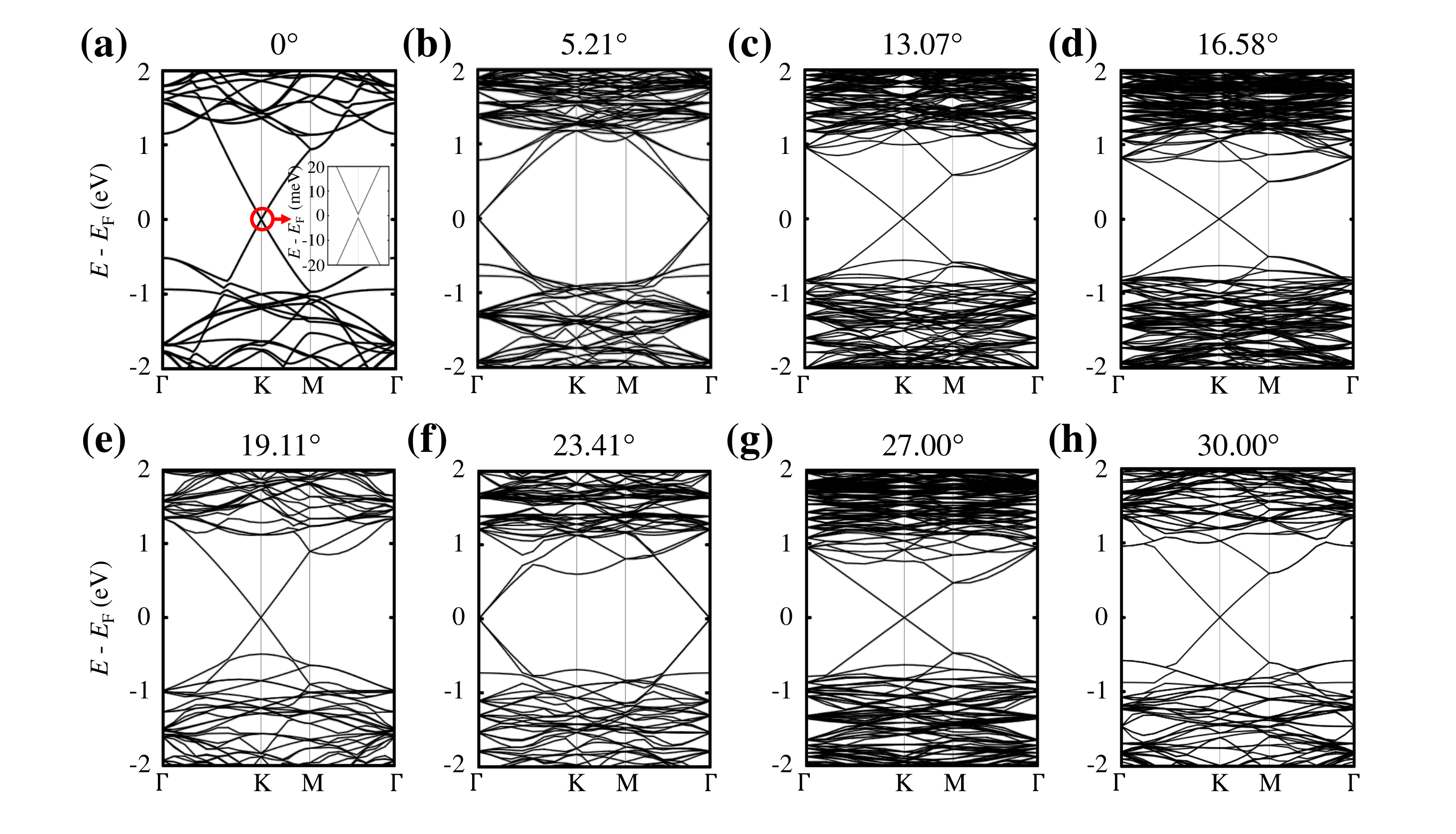}
\caption{(a--h) Electronic structure of twisted graphene/WSe$_{\rm{2}}$ heterostructures without spin-orbit interaction. In (a), inset shows small band gap opening at the Dirac point.}
\label{figs2}
\end{figure}
\clearpage
\newpage
\begin{figure}[p]
\includegraphics[width=1.0\columnwidth]{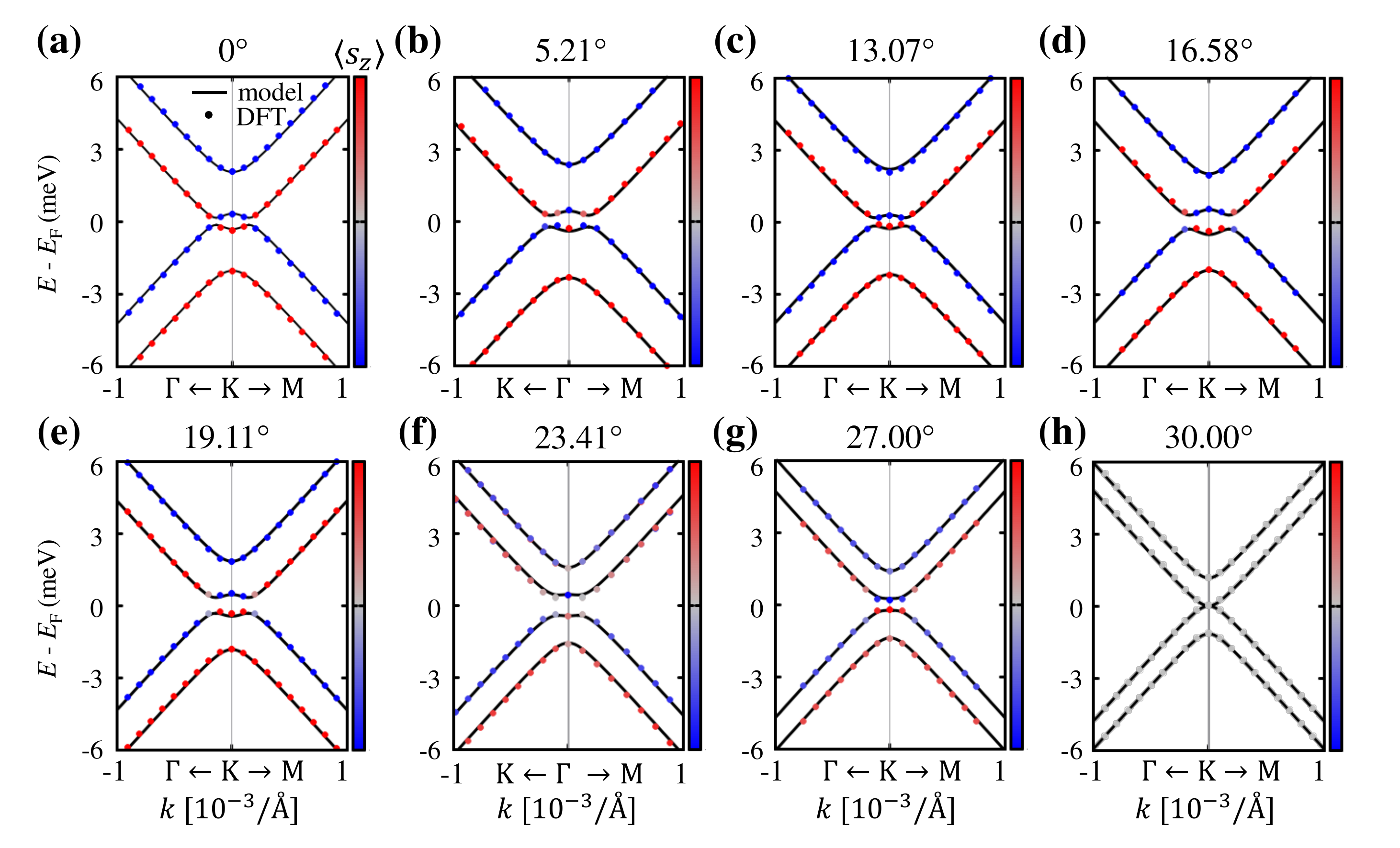}
\caption{(a--h) Spin-orbit proximitized carbon $p_z$ states of twisted graphene/WSe$_{\rm{2}}$ heterostructures. The colored dots show the energy eigenvalues calculated by DFT, which were fitted by the model defined as the Eq.~(1) in the main manuscript, plotted with black solid lines. In (a), colorbox indicates expectation value of out-of-plane spin components. }
\label{figs3}
\end{figure}
\clearpage
\newpage
\begin{figure}[p]
\includegraphics[width=1.0\columnwidth]{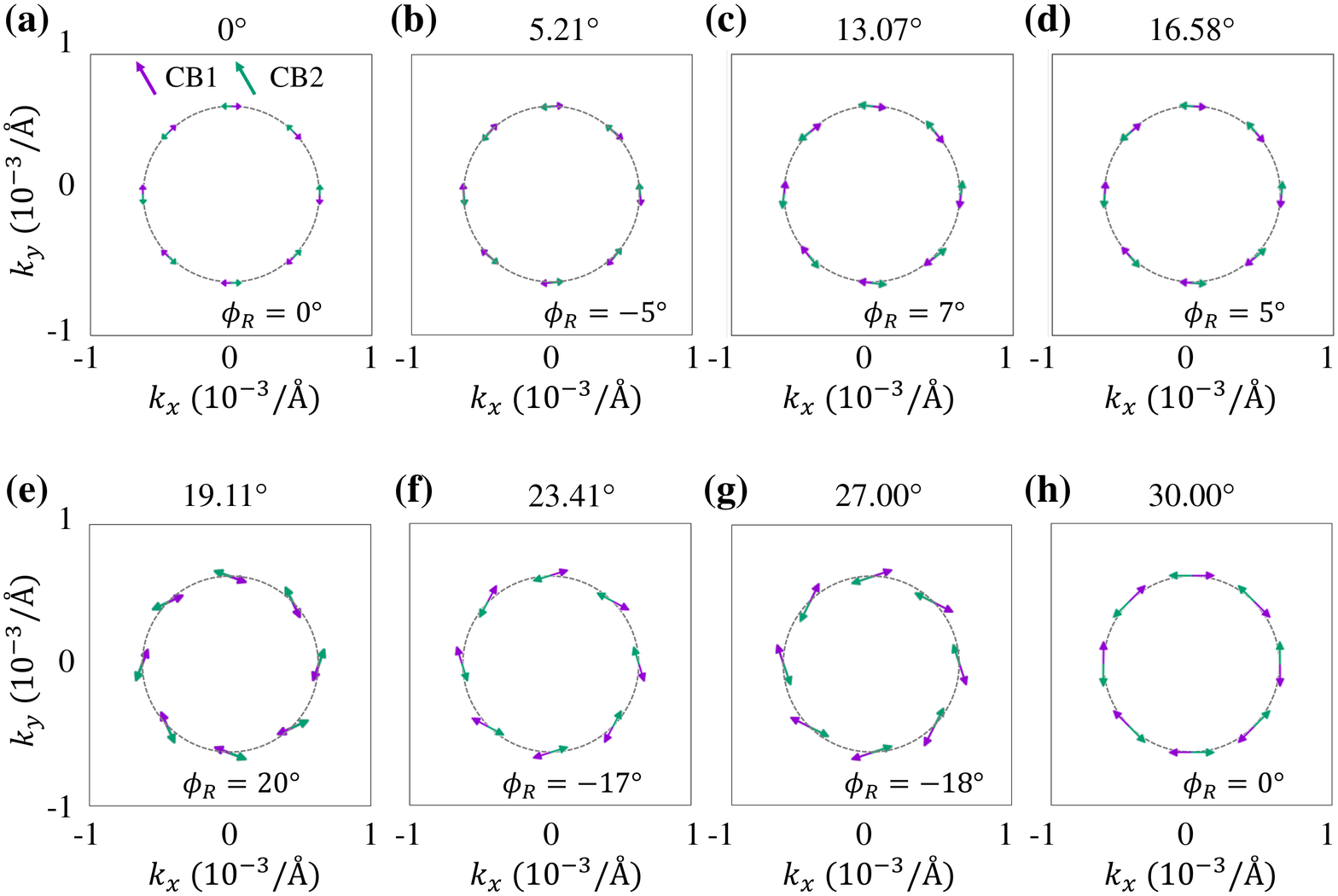}
\caption{
(a--h) In-plane spin angular momentum distributions of the heterostructures with twist angles from 0$^{\circ}$ to 30$^{\circ}$. The purple and cyan arrows indicate the (CB1) lowest and (CB2) second lowest conduction bands (two upper Dirac bands) with opposite spin-chiralities. The lowest and second highest valence bands (two lower Dirac bands) have exactly the same spin-chiralities with those of the CB1 and CB2, respectivley. 
In each figure, we represented Rashba phase angle $\phi_{R}$, which induces deviations from the ideal spin-momentum locking (highlighted by dashed circles). 
}
\label{figs4}
\end{figure}
\clearpage
\newpage
\begin{figure}[p]
\includegraphics[width=0.7\columnwidth]{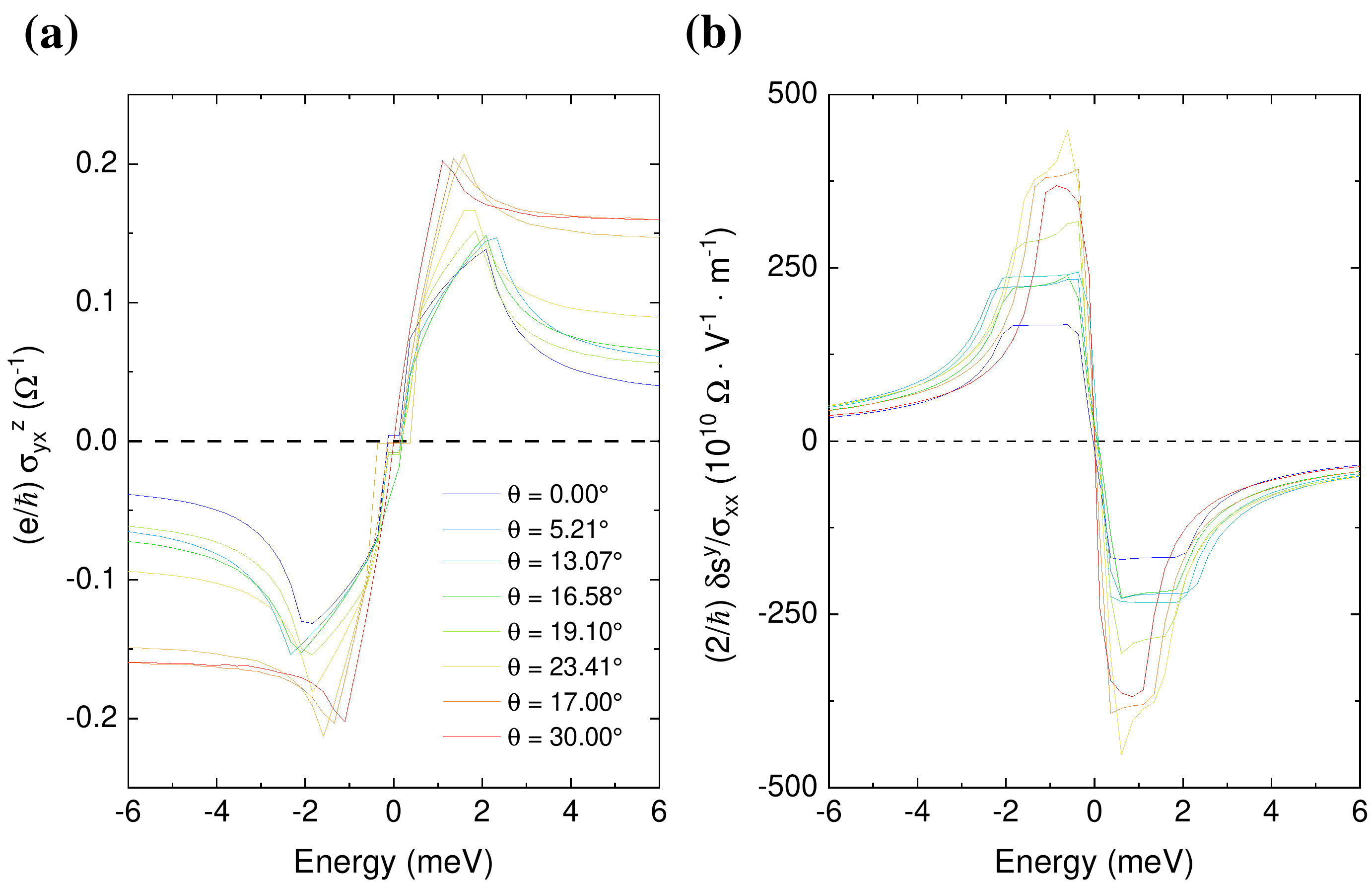}
\caption{Energy resolved (a) spin Hall conductivity and (b) Rashba-Edelstein efficiency at several twist angles. 
}
\label{figs6}
\end{figure}
\clearpage
\newpage
\begin{figure}[t]
\includegraphics[width=0.7\columnwidth]{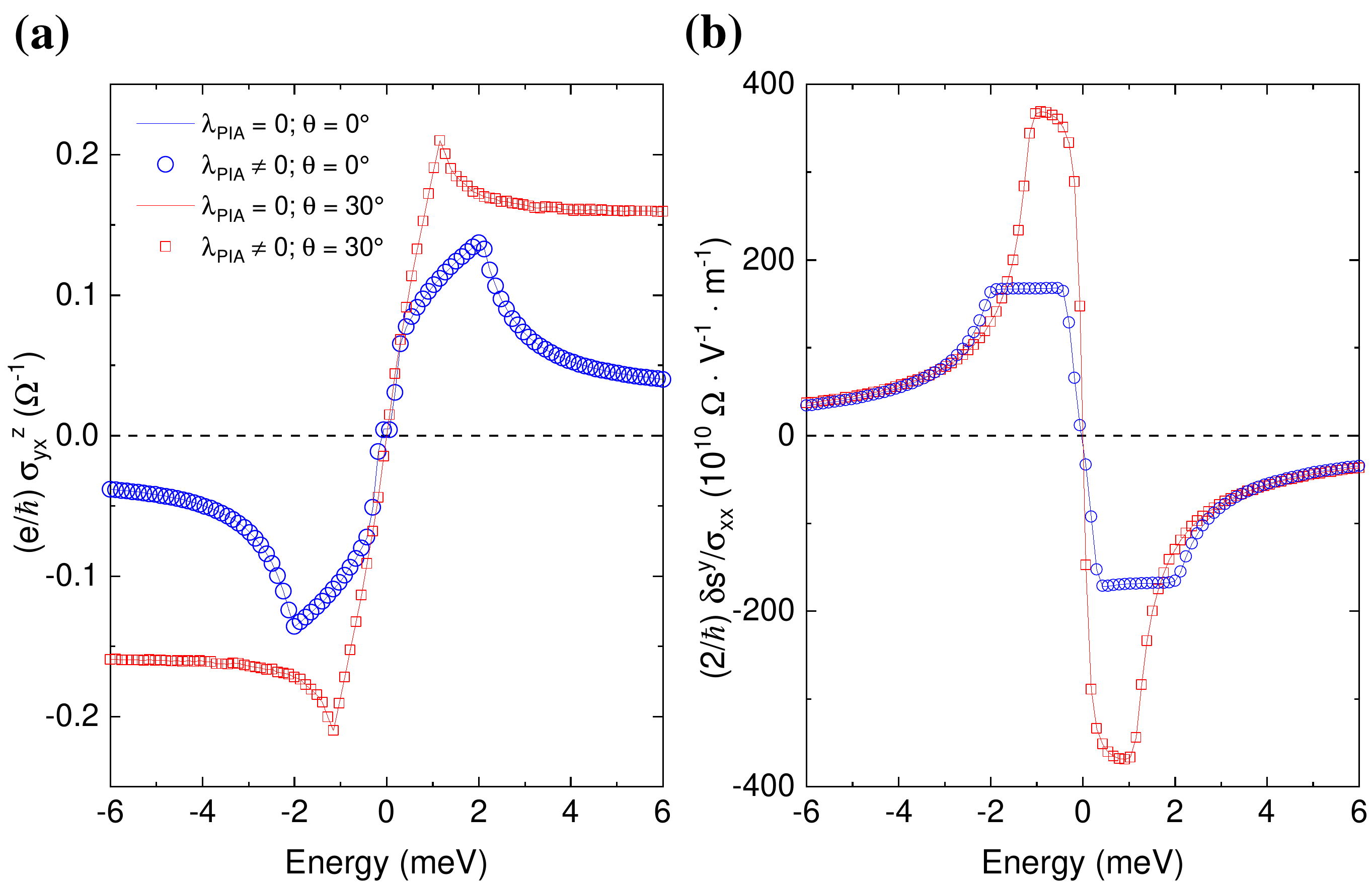}
\caption{Dependence of the (a) spin Hall conductivity and (b) Rashba-Edelstein efficiency with the pseudospin inversion asymmetry parameters $\lambda_{\rm{PIA}}^{A,B}$, represented collectively as $\lambda_{\rm{PIA}}$, at two sample twist angles. Here, $\lambda_{\rm{PIA}} \neq 0$ means the calculation was performed with full $\lambda_{\rm{PIA}}^{A,B}$ value as fitted from first-principles.
}
\label{figs7}
\end{figure}
\clearpage
\newpage
\begin{figure}[t]
\includegraphics[width=0.6\columnwidth]{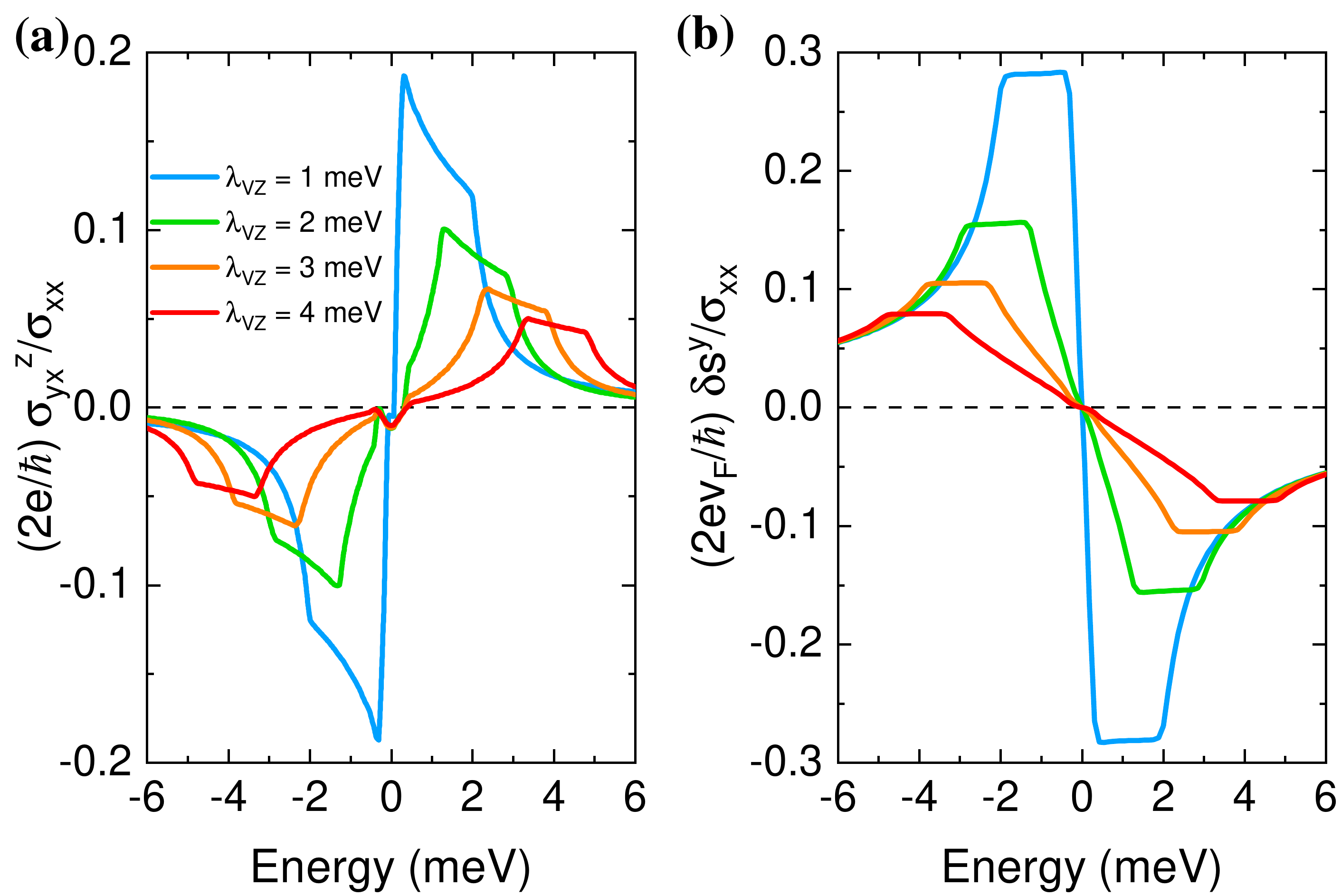}
\caption{Evolution of spin Hall and Rashba-Edelstein efficiencies with the valley Zeeman parameter. Here, all remaining parameters are fixed and assumed to coincide with those of the $0^{\circ}$ twisted structure.
}
\label{figs8}
\end{figure}

%
